\documentclass[a4paper,11pt,authordate1-4]{amsart}
\usepackage{authordate1-4}
\usepackage{latexsym}
\usepackage{amssymb}
\usepackage{amsfonts}
\usepackage{amsmath}
\usepackage{amsthm}
\usepackage{amscd}
\usepackage{enumerate}
\usepackage{url}
\usepackage{multirow}
\swapnumbers
\title[Mersenne Twister and Xorgens]{Combining the Mersenne Twister and the Xorgens designs}
\author{M. van de Vel}
\address{
        Faculty of sciences (FEW) \\
        VU Amsterdam \\
        Netherlands \\
	and \\
	Department of Mathematics \\
	University of Antwerp \\
	Belgium
        }
\date{\today}
\begin{document}
\addtolength{\hoffset}{-0.75cm}
\addtolength{\textwidth}{1.5cm}
\theoremstyle{plain}
\newtheorem{proposition}{Proposition}[section]
\newtheorem{theorem}[proposition]{Theorem}
\newtheorem{lemma}[proposition]{Lemma}
\newtheorem{corollary}[proposition]{Corollary}
\newtheorem{main}[proposition]{Main Theorem}
\theoremstyle{definition}
\newtheorem{definition}[proposition]{Definition}
\newtheorem{example}[proposition]{Example}
\newtheorem{examples}[proposition]{Examples}
\newtheorem{conclusion}[proposition]{Conclusion}
\newtheorem{remark}[proposition]{Remark}
\newtheorem{conventions}[proposition]{Conventions}
\newtheorem{problem}[proposition]{Problem}
\begin{abstract}
We combine the design of two \emph{random number generators},
\emph{Mersenne Twister} and \emph{Xorgens},
to obtain a new class of generators
with heavy-weight characteristic polynomials
(exceeded only by the {\sc well} generators)
and high speed (comparable with the originals).
Tables with parameter combinations are included
for state sizes ranging from 521 to 44497 bits
and each of the word lengths 32, 64, 128.
These generators passed all tests of the \emph{TestU01}-package
for each 32-bit integer part and each 64-bit derived real part of the output.
We determine \emph{dimension gaps}
for 32-bit words, neglecting the non-linear tempering,
and compare with an alternative experimental linear tempering.
\par
Categories and Subject Descriptors:
G.4 [{\bf Mathematical Software}]: Random number generator.
\par
General terms: Algorithms.
\par
Additional Key Words and Phrases:
Characteristic Polynomial,
Dimension Gap,
Mersenne Twister, Xorgens.
\end{abstract}
\maketitle
\section{Introduction}
\label{S:Intro}
\par
The \emph{Mersenne Twister} ({\sc mt}),
introduced by \cite{matsumo+nishimura:mersennetwister},
is one of the most popular \emph{random number generators} (RNGs) of the moment.
The bit size of the state space
is a (prime) exponent \( p \) of a \emph{Mersenne prime} \( 2^p - 1 \).
As \( p \) is not an integer multiple of the word size,
the state space contains an incomplete word
which requires a state transition ``twisting'' around the memory gap.
The period of this generator is the associated Mersenne prime.
\par
The original Mersenne Twisters had one of the sizes \( 11213, 19937 \),
each with a few variants.
The newer (and faster) SIMD oriented versions \cite{saito+matsumoto:SFMT_mersennetwister}
also exist in larger sizes.
More recently,
{\sc mt}-designs were presented by \cite{saito+matsumoto:gpu}
taking profit from parallel processing with graphic processors (GPUs).
\par
The \emph{Xorgens} ({\sc xg}) class
represents a rather different design of random number generators due to \cite{brent:xorgens}.
Building on earlier work of \cite{marsaglia:xorshifts},
Brent defines the core activity of {\sc xg}
in terms of \emph{bit shifts} and \emph{exclusive or} (\emph{xor}).
The state has \( 2^t \) bits (\( t \geq 8 \)).
The {\sc xg} class also includes generators of size \( 4224, 4480 \),
which are not a power of \( 2 \).
However, the size of an {\sc xg} is always an integer multiple of the word size.
\par
Except for these two,
the maximal possible period, \( 2^{2^{t}} - 1 \),
is a product of successive \emph{Fermat numbers} \( F_i , i = 1, \cdots t-1 \).
A proof that this period is attained
requires knowledge of its factorization,
which is available only for \( F_{i}, i \leq 11 \), \cite{brent:fermat}.
\cite{nandapalan+brent+murray+rendell} describe GPU processing
with the Xorgens design.
\par
Due to the involvement of Fermat number factorizations,
only ``small'' sizes are available for {\sc xg} generators.
An alternative design with Mersenne primes would give access to different and larger sizes and
--as it imposes a non-integral word size of the state---
one might expect variant behaviour through the ``twist'' of {\sc mt}.
This lead us to a sequence of well-behaved generators
with heavy-weight characteristic polynomials (CPs).
In fact,
the only RNGs known to us to have even heavier CPs belong to the {\sc well} class
(\cite{panneton+ecuyer+matsumoto}).
Our new RNGs have been labeled \emph{Mersenne Xorgens} ({\sc mxg})
to honour our two sources of inspiration.
\par
Table~\ref{S:Intro:T:speed_measurement} presents speed measurements
with a 2.40GHz Intel core.
We used mt19937ar, which is a 2002 revision of the original {\sc mt}
by T.~Nishimura and M.~Matsumoto, and more recent memt19937-II
by Shin Harase.
\par
We refer to \cite{mvdv+Ccode} for our C code used for speed measuring of 64-bit {\sc mxg}s.
Adaptation of the code to 32-bit words is straighforward.
The code for 128 bits requires a special header file (\texttt{emmintrin.h})
and some definitions of word operations like (arbitrary) shift and xorshift.
\par
Tables with good parameters
are presented in section \ref{S:MXGgenerators}.
The selected generators passed all Smallcrush, Crush, and BigCrush tests
in the package TestU01 \cite{ecuyer+simard:testu01},
applied to each 32-bit part.
\begin{table}[ht]
\begin{tabular}{ | l | r r || l | r r r |}
\hline
generator & 32 bit & 64 bit & generator & 32 bit & 64 bit & 128 bit \\
\hline
xg512        &  3.21s &  2.47s & mxg521   &  3.76s &  2.74s &  6.13s \\
xg1024       &  3.16s &  2.97s & mxg1279  &  3.56s &  3.65s &  6.70s \\
xg2048       &  1.56s &  3.31s & mxg2203  &  2.09s &  3.49s &  7.34s \\
xg4096       &  3.25s &  2.09s & mxg3217  &  3.87s &  3.51s &  6.61s \\
xg4224       &  1.56s &  2.18s & mxg4253  &  2.49s &  2.57s &  6.71s \\ 
xg4480       &  1.67s &  2.21s & mxg4423  &  2.68s &  3.59s &  7.54s \\
mt11213A     &  2.99s &  -     & mxg11213 &  3.46s &  3.80s &  7.16s \\
mt19937ar    &  2.96s &  2.94s & mxg19937 &  3.43s &  3.46s &  6.56s \\
memt19937-II &  5.62s &  -     & mxg21701 &  -     &  3.73s &  7.00s \\
well19937    &  7.42s &  -     & mxg23207 &  -     &  3.45s &  6.60s \\
well44497    & 12.98s &  -     & mxg44497 &  -     &  3.70s &  7.19s \\
% cmwc20632  & -      &  4.23s &          &        &        &        \\
\hline
\end{tabular}
\caption{Speed measurements: producing \( 10^9 \) integers}
\label{S:Intro:T:speed_measurement}
\end{table}
\par
{\sc well} generators have an additional property of \emph{maximal equidistribution} (ME),
and so does memt19937-II.
To measure deviation from ME,
we computed the (total) \emph{dimension gaps} \( \Delta \) (\cite{panneton+ecuyer+matsumoto})
for {\sc mxg} (dropping the non-linear Weyl tempering).
We treated each 32-bit word separately.
{\sc xg} does slightly better for smaller sizes and is somewhat worse for larger sizes.
On average, our scores are comparable with {\sc mt}s.
See section~2.
\par
However, for a fair judgement
we repeated our computations with an additional linear tempering
(two left and right xorshifts),
applied \emph{equally} to \emph{each} {\sc mxg} of \emph{any} word size.
The improvements on dimension gaps ranged between ``none'' and ``spectacular'':
four tempered 64-bit {\sc mxg}s deserve to be called \emph{near-ME} generators
(with all 32-bit words having \( \Delta < 10 \).
There is also one {\sc xg} like this.
This experiment suggests that simple, \emph{fashioned} linear temperings of {\sc mxg}
may produce other fast and near-ME generators.

\section{{\sc mxg} generators}
\label{S:MXGgenerators}
\subsection{Random number generators}
\label{S:MXGgenerators:D:general}
Let \( w \) denote a convenient word length (\( 32, 64, 128 \)) and
let \( n \) denote the number of \( w \)-bit words in the \emph{state} of a generator.
An eventual partial word is counted for one word
and its bits are thought of as being most significant (upper bits);
its lower bits are neglected.
The words of the state constitute an \( n \)-vector
\[
 \mathbf{x} := ( x_{n-1}, \ldots, x_{1}, x_{0} ),
\]
and its evolution is conceived as a continuation of the word sequence
with words \( x_n, x_{n+1}, \ldots \) and with a sliding window of \( n \) consecutive words.
Each step is perceived as a state transition \( \mathbf{x} \mapsto f(\mathbf{x}) \).
In practice,
one keeps running through the same memory positions by incrementing the word index modulo \( n \).
The output is given by a word-valued operator \( \mathbf{x} \mapsto o(\mathbf{x}) \).
One action cycle of the generator involves application of \( f \), then of \( o \).
\par
It is assumed here that at least \( f \) is linear.
Hence we deal with a \emph{linear recurrence}
having an associated \emph{characteristic polynomial} (CP).
For a state with \( p \) bits,
the maximal possible period is \( 2^p - 1 \) and is attained iff the CP is \emph{primitive},
\cite{lidl+niederreiter}.
\subsection{The Mersenne Twister}
\label{S:MXGgenerators:D:MersenneTwister}
We have a state space with  \( n \cdot w - r \) bits,
where \( 0 < r < w \).
Only the upper \( w - r \) bits of \( x_{0} \) are used.
The integer \( m < n \) (the \emph{step}) determines
how far to look ahead in the state space for a transition.
Consider the upper and lower masks
\[
  u^{w-r} := \underbrace{1 \ldots 1}_{w-r}\underbrace{0 \ldots 0}_{r}
  \qquad
  l^{r} := \underbrace{0 \ldots 0}_{w-r}\underbrace{1 \ldots 1}_{r}.
\]
The \( k \)-th state transformation of a Mersenne twister is given by
\[
  x_{k+n} := x_{k+m} \oplus ( x_k \cdot u^{w-r} \oplus x_{k+1} \cdot l^{r } ) M ,
  \qquad ( k \geq 0 ).
\]
The tokens '\( \oplus \)' and '\( \cdot \)' represent
bit-wise addition (``xor''), resp., multiplication (``and'').
The recurrence equation contains a \( w \times w \) bit matrix
\[
M :=
 \left(
 \begin{array}{ccccc}
  0 & 1 & 0 & \ldots & 0 \\
  0 & 0 & 1 & \ldots & 0 \\
  \vdots & \vdots & \vdots & \ddots & \vdots \\
  0 & 0 & 0 & \ldots & 1 \\
  a_{w-1} & a_{w-2} & \ldots & \ldots & a_{0}
 \end{array}
 \right)
\]
which does a right shift of a bit word \( ( b_{w-1}, \ldots, b_{0} ) \) by 1
and adds the word formed by the bottom row if \( b_{0} \not = 0 \).
The standard {\sc mt} recovers rather slowly from a near-zero state.
See \cite{panneton+ecuyer+matsumoto}
for a description of how the {\sc well} design avoids this phenomenon
(\emph{escaping zeroland}) because of a high CP weight.
The actual output of {\sc mt} is a word tempered with right- and left shifted bit-wise additions
(xorshifts).
See \cite{matsumo+nishimura:mersennetwister} for a precise description of the tempering.
\subsection{The Xorgens class}
\label{S:MXGgenerators:D:xorgen}
{\sc xg} generators \cite{brent:xorgens} have \( n \) entire words,
a step parameter \( m \) (as with {\sc mt}; Brent uses \( s := n - m \)), and
four parameters \( a, b, c, d \) controlling left and right shifts.
The \( k \)-th state transformation is
\[
  x_{k+n} := x_{k+m} (I \oplus C)(I \oplus D) \oplus x_k (I \oplus A)(I \oplus B),
  \qquad ( k \geq 0 ).
\]
The symbol \( I \) denotes the \( w \times w \) unit matrix.
In addition,
\( A, C \) are matrices reflecting left shifts by the respective amounts \( a, c \),
whereas \( B, D \) represent right shifts by the respective amounts \( b, d \).
\par
Brent prefers tempering with a \emph{Weyl generator},
whence the output becomes non-linear over the bit field.
This generator starts from any integer \( w_{0} \) that fits with the word length \( w \).
At the \( k \)th step the number \( w_{k} \) is generated
as \( w_{k-1} \) increased modulo \( 2^{w} \) with a \emph{Weyl constant};
a good choice is the odd integer approximation to \( 2^{w} \cdot (3 - \sqrt{5})/2 \).
The actual output of {\sc xg} is a \emph{numerical addition} modulo \( 2^w \),
\[
  x_{k+n} + ( w_{k} (I \oplus \mbox{\emph{W}}) ),
\]
where the matrix \emph{W} represents a right shift
over a fixed amount (the \emph{Weyl shift}).
This constant is chosen as
\( 16 \) (32-bit {\sc xg}) and
\( 27 \) (64-bit {\sc xg}).
\subsection{{\sc mxg} generators}
\label{S:MXGgenerators:D:mxg}
Random number generators of the {\sc mxg} class apply the Xorgens recurrence
to a state with the bit size of a Mersenne prime exponent.
Let \( r \) be the shortage of bits to have entire words.
We borrow from {\sc mt} design 
the idea of joining the upper \( w-r \) bits of a word
with the lower \( r \) bits of the next word.
With the above notations,
the recurrence becomes
\[
  x_{k+n} :=
  x_{k+m} (I \oplus C)(I \oplus D)
  \oplus
  ( x_k \cdot u^{w-r} \oplus x_{k+1} \cdot l^{r} ) (I \oplus A)(I \oplus B) .
\]
\par
We chose for tempering with the simple Weyl generator
to keep to our principle:
{\sc mxg} is simply {\sc xg} with the twist of {\sc mt}.
For the same reason,
we also copied the reliable {\sc xg} initialisation to {\sc mxg}.
% Note that this non-linear tempering may harm the equidistribution properties of {\sc mxg}
% (cf.~\ref{S:linfeedback:P:equidistribution}).
%
\subsection{The search process}
\label{S:MXGgenerators:D:searchprocess}
Having determined the generic transformations that run an {\sc mxg},
our next task is to assign parameters \( n, m, a, b, c, d \),
depending on the involved Mersenne prime,
to obtain a maximal period.
Brent's priorities to decide among multiple solutions
can be followed only to a certain extent.
\par
We consider pairs \( (a,b) \) and \( (c,d) \) which are relatively prime,
satisfy opposite strict inequalities,
and obey \( a + b \leq w \) and \( c + d \leq w \),
where \( w \) denotes the word length.
Left-right symmetry of pairs (as in {\sc xg}) is ruined by the ``twist'' design.
Our search proces counts down on the parameter
\( \delta := \min(a, b, c, d) \) from \( w/2 - 1 \) downto 6 (for 32-bit designs)
respectively, 17 (for 64-bit designs) and 40 (for 128-bit designs),
giving primary attention to high \( \delta \).
% Inside are two loops running through the list of admissable pairs
% to pick \( (a,b) \) and \( (c,d) \) with the appropriate minimum, \( \delta \).
\par
The list of admissible steps \( m \) reveals another difference with {\sc xg}.
Contrary to an observation of Brent for {\sc xg},
\( m \) (actually, \( s := n -  m \))
need not be relatively prime with the number of words, \( n \),
to obtain a primitive characteristic polynomial of an {\sc mxg}.
For instance,
Table~\ref{S:MXGgenerators:T:designs64bit} below lists a successful example of size 2281,
where \( n = 36 \) and \( m = 32 \).
As the involved words of a transition should not be neighbors
(which would interfere with the ``Mersenne twist'')
we let \( m \) range in \( 2 \ldots n - 1 \).
The resulting search space is roughly ten times the size of Brent's at comparable {\sc xg} sizes.
% 2 \cdot  69^2 \cdot (n - 2 ) & \mbox{(32-bit design)} \\
% 2 \cdot 159^2 \cdot (n - 2)  & \mbox{(64-bit design)} \\
% 2 \cdot 373^2 \cdot (n - 2)  & \mbox{(128-bit design)} \\
\par
The probability of a polynomial with a large prime degree \( p \) being primitive
is nearly exactly \( 1 / p \), \cite{lidl+niederreiter}.
Hence most of the search area is wasteland which should be crossed as fast as possible.
Unfortunately,
the efficient \emph{inversive-decimation} method of \cite{matsumo+nishimura:mersennetwister}
to decide on primitivity of the CP with \( O( p^2 ) \) operations
seems to be more complex for {\sc mxg}.
Refer to \ref{S:MXGgenerators:D:conclusion}.
\par
Contrasting with Brent's view,
we hold that the weight of the characteristic polynomial
should have more importance than just a final tie solver.
Although a high CP weight (optimally, near fifty percent) does not guarantee a good generator,
it does measure the intensity of mixing state bits.
Moreover,
a high weight facilitates recovery from near-zero states and
divergence of nearly identical states.
Therefore, it is a reasonable search directive.
We aim at CP weights ranging between \( \frac{1}{3} \) and \( \frac{2}{3} \) of the state bitsize.
\subsection{Description of the results}
\label{S:MXGgenerators:D:description}
\par
We searched (a fair portion of) the parameter space
for suitable parameter combinations with a heavy-weight primitive CP.
The corresponding generators survived all tests in the batteries
Small Crush (10),
Crush (96), and
Big Crush (106)
of TestU01, \cite{ecuyer+simard:testu01}.
As this package requires an RNG to produce 32-bit integers (if not reals),
we considered each generator
separately with each 32-bit word of its output.
In case of 128 bits,
we tested real numbers derived from the lower- and upper 64 bits separately.
Rare failures always showed a so-called ``p-value'' in between \( 10^{-4} \) and \( 10^{-3} \)
and did not reappear when the misbehaving test was performed three more times.
\par
We also computed the \emph{total dimension gap} \( \Delta \)
--a global measure for the deviation from \emph{maximal equidistribution},
\cite{panneton+ecuyer+matsumoto}--
for each 32-bit part of our generators and of all {\sc xg}s,
dropping the standard non-linear Weyl tempering.
Computing on 32-bit parts is not unusual, cf.~\cite{saito+matsumoto:SFMT_mersennetwister}.
For a fair judgement,
we also computed \( \Delta \) after applying a \emph{fixed} linear tempering,
\smallskip
\begin{verbatim}
      y ^= y>>1; y ^= y<<11;
      y ^= y>>7; y ^= y<<3;
\end{verbatim}
\smallskip
of an untempered word \( y \).
\subsubsection{The 32-bit case}
\label{S:MXGgenerators:D:32-bit_case}
We found generators with a primitive CP for Mersenne primes \( M_{13} \ldots M_{24} \).
See table \ref{S:MXGgenerators:T:designs32bit}.
Only at \( M_{13}, M_{14}, M_{15}, \) and \( M_{18} \)
our criterion on high CP weights is met,
yet the weights are much higher than those of nearby 32-bit {\sc xg} sizes
(see \cite{brent:xorgens} or
\url{http://maths-people.anu.edu.au/~brent/ftp/random/xortable.txt}):

\smallskip
\begin{tabular}{ l l l l}
& {\sc xg}512: 185; & {\sc xg}1024: 225; & {\sc xg}2048: 213; \\
& {\sc xg}4096: 251; & {\sc xg}4224: 243; & {\sc xg}4480: 251. \\
\end{tabular}
\smallskip

For 32-bit {\sc mt} of size 11213 we found \( \Delta = 4087 \)
whereas the standard 32-bit {\sc mt} of size 19937 has \( \Delta = 6750 \)
\cite{saito+matsumoto:SFMT_mersennetwister}.
These values are more than twice our tempered {\sc mxg} values.
For 32-bit {\sc xg} generators we found the following \( \Delta \)-values
(untempered,tempered):

\smallskip
\begin{tabular}{ l l l l}
& {\sc xg}512: (142,92); & {\sc xg}1024: (141,100); & {\sc xg}2048: (465,335); \\
& {\sc xg}4096: (845,454); & {\sc xg}4224: (1838,622); & {\sc xg}4480: (2038,590). \\
\end{tabular}

% \begin{table}[ht]
% \caption{Summary of search activities: 32-bit case}
% \begin{tabular}{| l | r r r r | }
% \hline
% Mersenne & size    & size of      &          & successful \\
% prime    & in bits & search space & searched & parameters \\
% \hline
% \( M_{13} \) &   521 &   133,308 &  40,000 & 111 \\
% \( M_{14} \) &   607 &   152,352 & 100,000 & 239 \\
% \( M_{15} \) &  1279 &   352,314 & 100,000 & 137 \\
% \( M_{16} \) &  2203 &   628,452 & 100,000 &  58 \\
% \( M_{17} \) &  2281 &   657,018 & 100,000 &  82 \\
% \( M_{18} \) &  3217 &   933,156 & 100,000 &  43 \\
% \( M_{19} \) &  4253 & 1,247,382 & 520,000 & 216 \\
% \( M_{20} \) &  4423 & 1,304,514 &  90,000 &  29 \\
% \( M_{21} \) &  9689 & 2,856,600 &  60,000 &   8 \\
% \( M_{22} \) &  9941 & 2,942,298 &  26,000 &   3 \\
% \( M_{23} \) & 11213 & 3,323,178 &  50,000 &   5 \\
% \( M_{24} \) & 19937 & 5,913,162 &  55,200 &   5 \\
% \( M_{25} \) & 21701 & 6,465,438 &   3,250 &   0 \\
% \hline
% \end{tabular}
% \label{S:MXGgenerators:T:search_activities32bit}
% \end{table}
%
%
\begin{table}[ht]
\begin{tabular}{ | l | r r r | r r r r r | r | r r | }
\hline
& \multirow{2}{*}{bits} & \multirow{2}{*}{n} & \multirow{2}{*}{r} &
\multirow{2}{*}{m} & \multirow{2}{*}{a} & \multirow{2}{*}{b} & \multirow{2}{*}{c} & \multirow{2}{*}{d} & \multirow{2}{*}{weight} & \multicolumn{2}{c|}{$\Delta$} \\
 & & & & & & & & & & untemp & temp \\
\hline  
\( M_{13} \) &   521 &  17 & 23 &  10 & 11 & 15 & 14 & 11 &  261 &  150 &   38 \\
\( M_{14} \) &   607 &  19 &  1 &   3 & 17 & 13 &  7 & 22 &  303 &  167 &   76 \\
\( M_{15} \) &  1279 &  40 &  1 &  26 & 13 & 10 &  9 & 23 &  513 &  360 &  117 \\
\( M_{16} \) &  2203 &  69 &  5 &  16 & 16 & 13 & 10 & 11 &  855 &  433 &  206 \\
\( M_{17} \) &  2281 &  72 & 23 &  65 & 13 & 18 & 15 & 14 &  923 &  983 &  298 \\
\( M_{18} \) &  3217 & 101 & 15 &  95 & 19 & 13 & 15 & 16 & 1203 & 1040 &  520 \\
\( M_{19} \) &  4253 & 133 &  3 &  31 & 11 &  8 &  9 & 16 & 1045 & 1497 &  418 \\
\( M_{20} \) &  4423 & 139 & 25 &  79 & 15 & 14 & 11 & 18 & 1383 & 1157 &  363 \\
\( M_{21} \) &  9689 & 302 &  7 & 295 & 14 & 13 & 13 & 18 & 1647 & 4575 & 2077 \\
\( M_{22} \) &  9941 & 311 & 11 &  17 & 13 & 14 & 17 & 14 & 1765 & 3719 & 1799 \\
\( M_{23} \) & 11213 & 351 & 19 & 330 & 17 & 13 & 15 & 17 & 2021 & 5363 & 1883 \\
\( M_{24} \) & 19937 & 621 & 31 & 319 & 13 & 19 & 16 & 15 &  581 & 8047 & 2488 \\
\hline
\end{tabular}
\caption{{\sc mxg} designs (32-bit words)}
\label{S:MXGgenerators:T:designs32bit}
\end{table}
%
% \begin{table}[ht]
% \caption{Summary of search activities: 64-bit case}
% \begin{tabular}{| l | r r r r | }
% \hline
% Mersenne & size    & size of      &          & successful \\
% prime    & in bits & search space & searched & parameters \\
% \hline
% \( M_{13} \) &    521 &    303,372 & 200,000 & 539 \\
% \( M_{14} \) &    607 &    353,934 & 200,000 & 472 \\
% \( M_{15} \) &  1,279 &    859,554 & 200,000 & 257 \\
% \( M_{16} \) &  2,203 &  1,777,670 & 300,000 & 214 \\
% \( M_{17} \) &  2,281 &  1,820,232 & 950,000 & 477 \\
% \( M_{18} \) &  3,217 &  2,426,976 & 750,000 & 317 \\
% \( M_{19} \) &  4,253 &  3,235,968 & 320,000 & 123 \\
% \( M_{20} \) &  4,423 &  3,387,654 & 270,000 & 102 \\
% \( M_{21} \) &  9,689 &  7,533,738 & 100,000 &  12 \\
% \( M_{22} \) &  9,941 &  7,735,986 &  80,000 &  12 \\
% \( M_{23} \) & 11,213 &  8,747,226 & 115,000 &  13 \\
% \( M_{24} \) & 19,937 & 15,623,658 &  58,000 &   8 \\
% \( M_{25} \) & 21,701 & 17,039,394 & 310,000 &  17 \\ 
% \( M_{26} \) & 23,207 & 18,354,006 &  48,000 &   4 \\ 
% \( M_{27} \) & 44,497 & 35,191,152 &  32,000 &   2 \\ 
% \hline
% \end{tabular}
% \label{S:MXGgenerators:T:search_activities64bit}
% \end{table}
% %
\begin{table}[ht]
\fontsize{9}{11}\selectfont
\begin{tabular}{ | l | r r r | r r r r r | r | r r | r r | }
\hline
& \multirow{3}{*}{bits} & \multirow{3}{*}{n} & \multirow{3}{*}{r} &
\multirow{3}{*}{m} & \multirow{3}{*}{a} & \multirow{3}{*}{b} & \multirow{3}{*}{c} & \multirow{3}{*}{d} & \multirow{3}{*}{weight} & \multicolumn{4}{c|}{$\Delta$} \\
\cline{11-14}
 & & & & & & & & & & \multicolumn{2}{c|}{word 1} & \multicolumn{2}{c|}{word 0} \\
 & & & & & & & & & & unt. & temp. & unt. & temp. \\
\hline
\( M_{13} \) &   521 &   9 & 55 &   4 & 32 & 27 & 28 & 33 &   261 &   146 & 2 &  184 & 5 \\
\( M_{14} \) &   607 &  10 & 33 &   6 & 31 & 26 & 27 & 34 &   303 &   126 & 0 &  133 & 1 \\
\( M_{15} \) &  1279 &  20 &  1 &   6 & 27 & 32 & 33 & 29 &   639 &   320 & 4 &  685 & 10 \\
\( M_{16} \) &  2203 &  35 & 37 &  23 & 23 & 29 & 25 & 22 &  1089 &   335 & 2 &  821 & 8 \\
\( M_{17} \) &  2281 &  36 & 23 &  23 & 25 & 19 & 19 & 23 &  1121 &   305 & 3 &  314 & 11 \\
\( M_{18} \) &  3217 &  51 & 47 &  29 & 22 & 35 & 37 & 21 &  1519 &   108 & 3 &  225 & 6 \\
\( M_{19} \) &  4253 &  67 & 35 &   8 & 25 & 26 & 25 & 23 &  1983 &  3648 & 80 & 3935 & 52 \\
\( M_{20} \) &  4423 &  70 & 57 &  62 & 31 & 28 & 23 & 34 &  2057 &   378 & 10 & 1038 & 22 \\
\( M_{21} \) &  9689 & 152 & 39 &  41 & 29 & 31 & 29 & 28 &  3925 & 11700 & 5403 & 12041 & 229 \\
\( M_{22} \) &  9941 & 156 & 43 & 110 & 29 & 32 & 28 & 27 &  4013 &  6793 & 342 & 7906 & 18 \\
\( M_{23} \) & 11213 & 176 & 51 &  93 & 27 & 34 & 31 & 28 &  4355 &  86 & 61 & 5010 & 69 \\
\( M_{24} \) & 19937 & 312 & 31 & 275 & 35 & 29 & 28 & 35 &  6913 &  5687 & 177 & 5863 & 154 \\
\( M_{25} \) & 21701 & 339 & 59 & 308 & 33 & 29 & 27 & 37 &  5765 &  5303 & 231 & 5304 & 293 \\
\( M_{26} \) & 23209 & 363 & 23 & 229 & 31 & 29 & 31 & 32 &  7853 & 30598 & 257 & 30597 & 935 \\
\( M_{27} \) & 44497 & 696 & 47 & 662 & 31 & 33 & 31 & 29 & 11663 & 49289 & 1462 & 49984 & 1971 \\
\hline
\end{tabular}
\caption{{\sc mxg} designs (64-bit words)}
\label{S:MXGgenerators:T:designs64bit}
\end{table}
\subsubsection{The 64-bit case}
\label{S:MXGgenerators:D:64-bit_case}
The parameters given in table \ref{S:MXGgenerators:T:designs64bit}
for \( M_{25} \) and \( M_{27} \) have the highest CP weight found,
but they do not meet our ``1/3'' criterion.
Here are the corresponding Xorgens CP weights:

\smallskip
\begin{tabular}{ l l l l}
& {\sc xg}512: 231; & {\sc xg}1024: 439; & {\sc xg}2048: 745; \\
& {\sc xg}4096: 961; & {\sc xg}4224: 987; & {\sc xg}4480: 951. \\
\end{tabular}
\smallskip

The total equidistribution deficits of 64-bit {\sc mxg} are given in the last columns,
together with the values in case of a fixed linear tempering.
The 64-bit {\sc mt}19937 has \( \Delta \)(word1,word0)\( =  (4161,10299) \)
As to Xorgens,
the values of \( \Delta \)(word1,word0), untempered/tempered, are:

\smallskip
\begin{tabular}{ l l l }
& {\sc xg}512: (74/7, 123/4); & {\sc xg}1024: 105/4, 350/11); \\
& {\sc xg}2048: (209/100, 416/102); & {\sc xg}4096: (666/26, 1715/50); \\
& {\sc xg}4224: (1202/48, 2334/70); & {\sc xg}4480: (2710/634, 3351/949). \\
\end{tabular}
%\newline
%
\subsubsection{The 128-bit case}
\label{S:MXGgenerators:D:128-bit_case}
%
%
% RNG design tends towards larger sizes: not just states, but also word lengths.
As xorshift operators involve more bits with longer words,
state transitions are likely to gain complexity.
Our search process confirms that
both the \emph{average} and \emph{maximal} weight of primitive polynomials grow significantly
with the wordsize.
Therefore,
we extrapolated the {\sc mxg}-design to 128 bit,
using SIMD (Single Instruction Multiple Data) to handle \( 128 \)-bit words.
Table \ref{S:MXGgenerators:T:designs128bit} presents good parameters.
For each target state size,
we achieved our criterion on high CP weights rather easily.
\cite{saito+matsumoto:SFMT_mersennetwister} report that
the CP weight of 128-bit sfmt19937 is 6711 and that
the 32-bit well19937 has CP weight 8585.
\par
Table \ref{S:MXGgenerators:T:dimgap128bit} shows
the dimension gaps of 128-bit generators for each 32-bit word.
For sfmt19937,
\( \Delta \) equals 14089 (64-bit) or 28676 (128-bit).
%
% \begin{table}[ht]
% \caption{Summary of search activities: 128-bit case}
% \begin{tabular}{| l | r r r r | }
% \hline
% Mersenne & size    & size of      &          & successful \\
% prime    & in bits & search space & searched & parameters \\
% \hline
% \( M_{13} \) &    521 &    417,387 &   6,000 &  14 \\
% \( M_{14} \) &    607 &    417,387 &   8,000 &  25 \\
% \( M_{15} \) &  1,279 &  1,113,032 &  25,000 &  20 \\
% \( M_{16} \) &  2,203 &  2,226,064 &  25,000 &  17 \\
% \( M_{17} \) &  2,281 &  2,226,064 &  30,000 &  16 \\
% \( M_{18} \) &  3,217 &  3,339,096 &  50,000 &  16 \\
% \( M_{19} \) &  4,253 &  4,452,128 &  40,000 &  18 \\
% \( M_{20} \) &  4,423 &  4,591,257 &  40,000 &   9 \\
% \( M_{21} \) &  9,689 & 10,295,546 &  25,000 &   3 \\
% \( M_{22} \) &  9,941 & 10,573,804 &  80,000 &   7 \\
% \( M_{23} \) & 11,213 & 11,965,094 &  50,000 &   6 \\
% \( M_{24} \) & 19,937 & 21,425,866 &  36,000 &   1 \\
% \( M_{25} \) & 21,701 & 23,373,672 &  28,000 &   3 \\ 
% \( M_{26} \) & 23,209 & 25,043,220 &  12,000 &   1 \\ 
% \( M_{27} \) & 44,497 & 48,138,634 &   7,500 &   1 \\ 
% \hline
% \end{tabular}
% \label{S:MXGgenerators:T:search_activities128bit}
% \end{table}
%
\begin{table}[ht]
\begin{tabular}{ | l | r r r | r r r r r | c | }
\hline
             &  bits &   n &   r &   m &  a &  b &  c &  d & weight \\
\hline
\( M_{13} \) &   521 &   5 & 119 &   3 & 57 & 68 & 65 & 57 &   261 \\
\( M_{14} \) &   607 &   5 &  33 &   3 & 57 & 61 & 57 & 56 &   303 \\
\( M_{15} \) &  1279 &  10 &   1 &   3 & 56 & 71 & 69 & 58 &   639 \\
\( M_{16} \) &  2203 &  18 & 101 &   7 & 59 & 60 & 64 & 61 &  1103 \\
\( M_{17} \) &  2281 &  18 &  23 &   3 & 60 & 67 & 67 & 59 &  1145 \\
\( M_{18} \) &  3217 &  26 & 111 &  18 & 69 & 59 & 61 & 66 &  1597 \\
\( M_{19} \) &  4253 &  34 &  99 &  26 & 64 & 63 & 58 & 69 &  2097 \\
\( M_{20} \) &  4423 &  35 &  57 &  29 & 61 & 60 & 59 & 63 &  2163 \\
\( M_{21} \) &  9689 &  76 &  39 &  69 & 62 & 65 & 63 & 62 &  4621 \\
\( M_{22} \) &  9941 &  78 &  43 &  11 & 61 & 62 & 59 & 58 &  4681 \\
\( M_{23} \) & 11213 &  88 &  51 &  28 & 61 & 63 & 64 & 61 &  5163 \\
\( M_{24} \) & 19937 & 156 &  31 &  85 & 61 & 67 & 64 & 63 &  8823 \\
\( M_{25} \) & 21701 & 170 &  59 & 133 & 63 & 64 & 62 & 61 &  9785 \\
\( M_{26} \) & 23209 & 182 &  87 &  99 & 64 & 63 & 61 & 67 &  9965 \\
\( M_{27} \) & 44497 & 348 &  47 & 235 & 62 & 63 & 65 & 63 & 17293 \\
\hline
\end{tabular}
\caption{{\sc mxg} designs (128-bit words)}
\label{S:MXGgenerators:T:designs128bit}
\end{table}
%
% François Panneton, Pierre L’Ecuyer
% Resolution-stationary random number generators
% Mathematics and Computers in Simulation , vol. 80, no. 6, pp. 1096-1103, 2010
% DOI: 10.1016/j.matcom.2007.09.014
%
\begin{table}[ht]
\fontsize{9}{11}\selectfont
\begin{tabular}{ | l | c | r r | r r | r r | r r | }
\hline
 &  bits & \multicolumn{8}{c|}{$\Delta$} \\
\cline{3-10}
 &       & \multicolumn{2}{c|}{word 3} & \multicolumn{2}{c|}{word 2} & \multicolumn{2}{c|}{word 1} & \multicolumn{2}{c|}{word 0} \\
 & & unt. & temp. & unt. & temp. & unt. & temp. & unt. & temp. \\
\hline
\( M_{13} \) &   521 &   132 &    60 &   157 &     2 &   148 &    32 &   160 &     4 \\
\( M_{14} \) &   607 &   668 &   347 &   813 &    64 &   669 &   274 &   821 &    57 \\
\( M_{15} \) &  1279 &   114 &    87 &    39 &     2 &   139 &   114 &   140 &     5 \\
\( M_{16} \) &  2203 &  2038 &   477 &  2646 &  2366 &  2463 &   124 &  2647 &  2645 \\
\( M_{17} \) &  2281 &   969 &   872 &   896 &     6 &  1076 &   619 &  1089 &    78 \\
\( M_{18} \) &  3217 &  1365 &    44 &  1526 &     2 &  1511 &     5 &  1526 &  1527 \\
\( M_{19} \) &  4253 &  2456 &  2052 &  2734 &     2 &  2714 &    18 &  2733 &  2732 \\
\( M_{20} \) &  4423 &  5083 &  2577 &  5976 &  5975 &  5609 &  1806 &  5976 &  5977 \\
\( M_{21} \) &  9689 & 14527 & 10927 & 14894 & 11646 & 14872 &  7612 & 14894 & 13113 \\
\( M_{22} \) &  9941 & 10478 &  2796 & 11950 & 11950 & 11465 &   517 & 11950 & 11949 \\
\( M_{23} \) & 11213 & 14103 &  5827 & 15165 & 15165 & 14595 &  3082 & 15165 & 11166 \\
\( M_{24} \) & 19937 & 21411 & 17941 & 20932 &  9612 & 21414 & 12799 & 21436 & 10106  \\
\( M_{25} \) & 21701 & 32050 & 21218 & 33362 & 33362 & 33359 & 15562 & 33362 & 33361 \\
\( M_{26} \) & 23209 & 24930 & 20222 & 24954 &  2207 & 24933 & 14905 & 24951 &  4359 \\
\( M_{27} \) & 44497 & 75356 & 38085 & 78848 & 78846 & 77439 & 27083 & 78847 & 78846 \\
\hline
\end{tabular}
\caption{128-bit Deficit, untempered/tempered}
\label{S:MXGgenerators:T:dimgap128bit}
\end{table}
% Data set on Research gate: DOI: 10.13140/2.1.1185.7924
% https://www.researchgate.net/publication/267439134_mxg_a_new_clas_of_random_number_generators_in_between_Mersenne_Twister_and_Xorgens
%
\subsection{Conclusion}
\label{S:MXGgenerators:D:conclusion}
{\sc mxg} generators combine the Xorgens and Mersenne Twister designs,
have heavy CPs, and pass all standard statistical tests.
Replacing the nonlinear Weyl tempering by
a fixed linear tempering provides several near ME generators.
Fashioned tempering, especially in the range of 64-bit generators,
may produce more of these.
It is an open problem whether, for an {\sc mxg} of bitsize \( p \),
primitivity of the CP can be decided with at most \( O(p^2) \) operations
as is the case with {\sc mt} \cite{matsumo+nishimura:mersennetwister}.

\ \newline
\bibliography{mxg}
\bibliographystyle{authordate1}
%\bibliographystyle{ws-book-har}
% standards: abbrv, alpha, plain, or unsrt
% latex myarticle
% bibtex myarticle
% latex myarticle
% latex myarticle
% \input{sc3}
%
% \input{connections}
%
\end{document}